\documentclass[
    ,final            
  ]
  {aipproc}

\layoutstyle{6x9}
\usepackage{amssymb}
\usepackage{amsmath}

\providecommand{\PYTHIA}{{\sc pythia }}
\providecommand{\HERWIG}{{\sc herwig }}
\def\pTa{p_{T,\,1}}
\def\pTb{p_{T,\,2}}
\def\mean#1{\ensuremath{\left<#1\right>}}

\begin{document}

\title{Low-$x$ QCD studies with forward jets in proton-proton collisions at $\sqrt{s}$~=~14~TeV}
\classification{12.38.-t, 12.38.Qk, 13.87.-a}

\keywords{Forward jets, jet algorithms, low-$x$ QCD, BFKL, Mueller-Navelet dijets, LHC.}
\author{Salim Cerci}{address={Cukurova University, Fen-Ed. Fakultesi 01330, Adana, Turkey}}
\author{David~d'Enterria$^\dagger$ for the CMS collaboration}{address={CERN, CH-1211 Geneva 23, Switzerland} }

\begin{abstract}
Forward (di)jet measurements are a useful tool to constrain the Parton Distribution Functions (PDFs) 
at low values of parton momentum fraction $x$ 
and to study the possible onset of BFKL or gluon saturation QCD evolutions in the proton. We present 
studies of jet reconstruction capabilities in the CMS Hadron Forward (HF) calorimeter (3$<|\eta|<$5).
The expected sensitivity of the inclusive forward jet $p_T$ spectrum to the proton PDF, as well 
as the azimuthal decorrelation of Mueller-Navelet jets with a large 
rapidity separation are presented for p-p collisions at $\sqrt{s}$~=~14~TeV.
\end{abstract}

\maketitle


\section{Introduction}
\label{sec:intro}
Studies of the high-energy (low-$x$) limit of Quantum Chromodynamics (QCD) have attracted 
much theoretical interest in the last 10--15 years, in the context of deep-inelastic (DIS) electron-proton as well as 
of nucleus-nucleus collisions~\cite{cgc}. For decreasing parton momentum fraction 
$x$~=~$p_{\mbox{\tiny{\it parton}}}/p_{\mbox{\tiny{\it hadron}}}$, the gluon density is 
observed to grow rapidly in the hadronic wave-functions. As long as the densities are not too high, 
this growth is described by the Dokshitzer-Gribov-Lipatov-Altarelli-Parisi (DGLAP)~\cite{dglap} 
or by the Balitski-Fadin-Kuraev-Lipatov (BFKL)~\cite{bfkl} evolution equations which govern, 
respectively, parton radiation in $Q^2$ and $x$.
Eventually, at high enough centre-of-mass energies (i.e. at very small $x$) the gluon density will 
be so large that non-linear (gluon-gluon fusion) effects will become important, saturating the 
growth of the parton densities~\cite{cgc}.\\ 

\begin{figure}[htb]
\centering
\includegraphics[width=0.42\textwidth,height=5.cm]{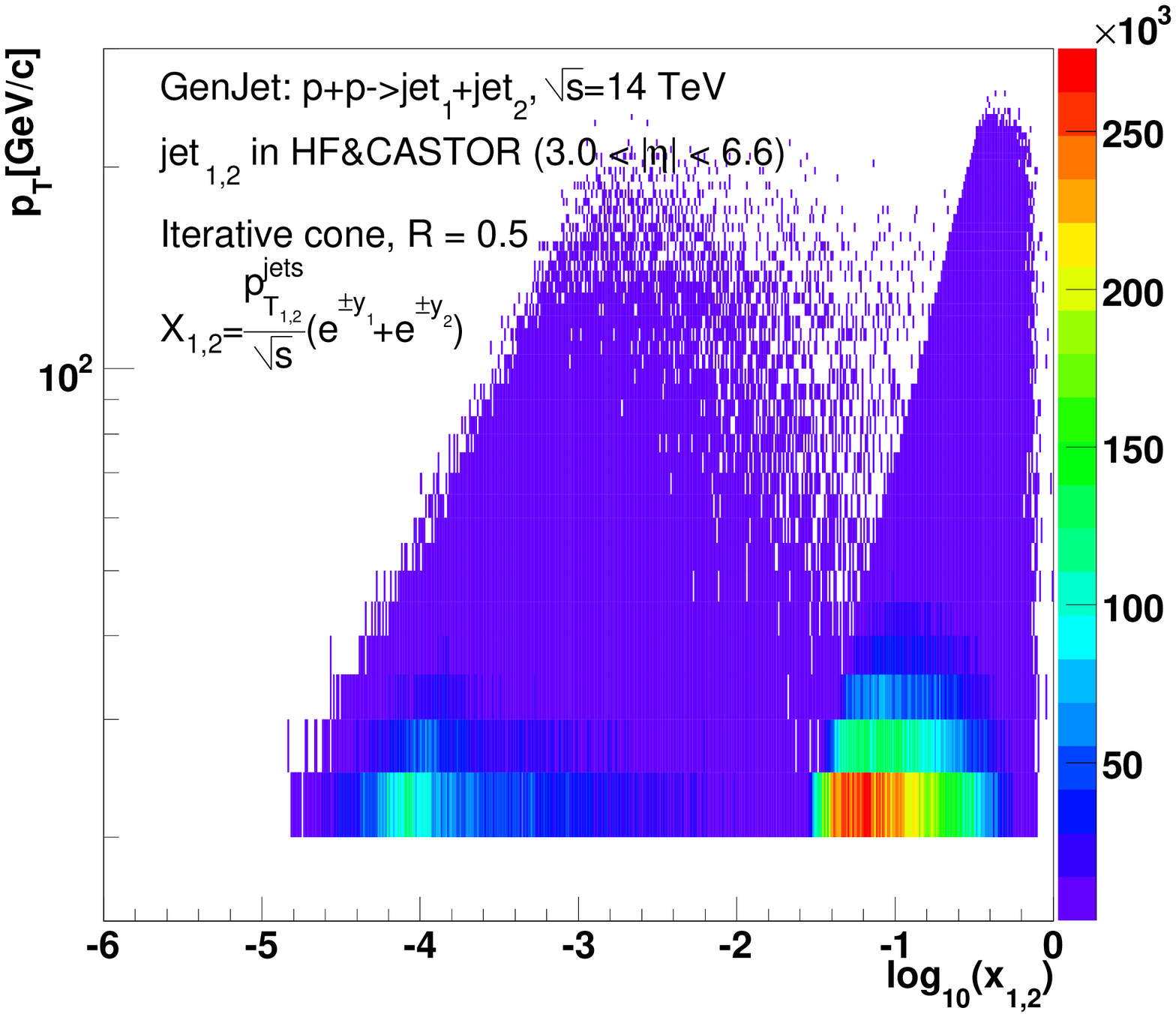}\hspace{0.2cm}
\includegraphics[width=0.58\textwidth,height=5.cm]{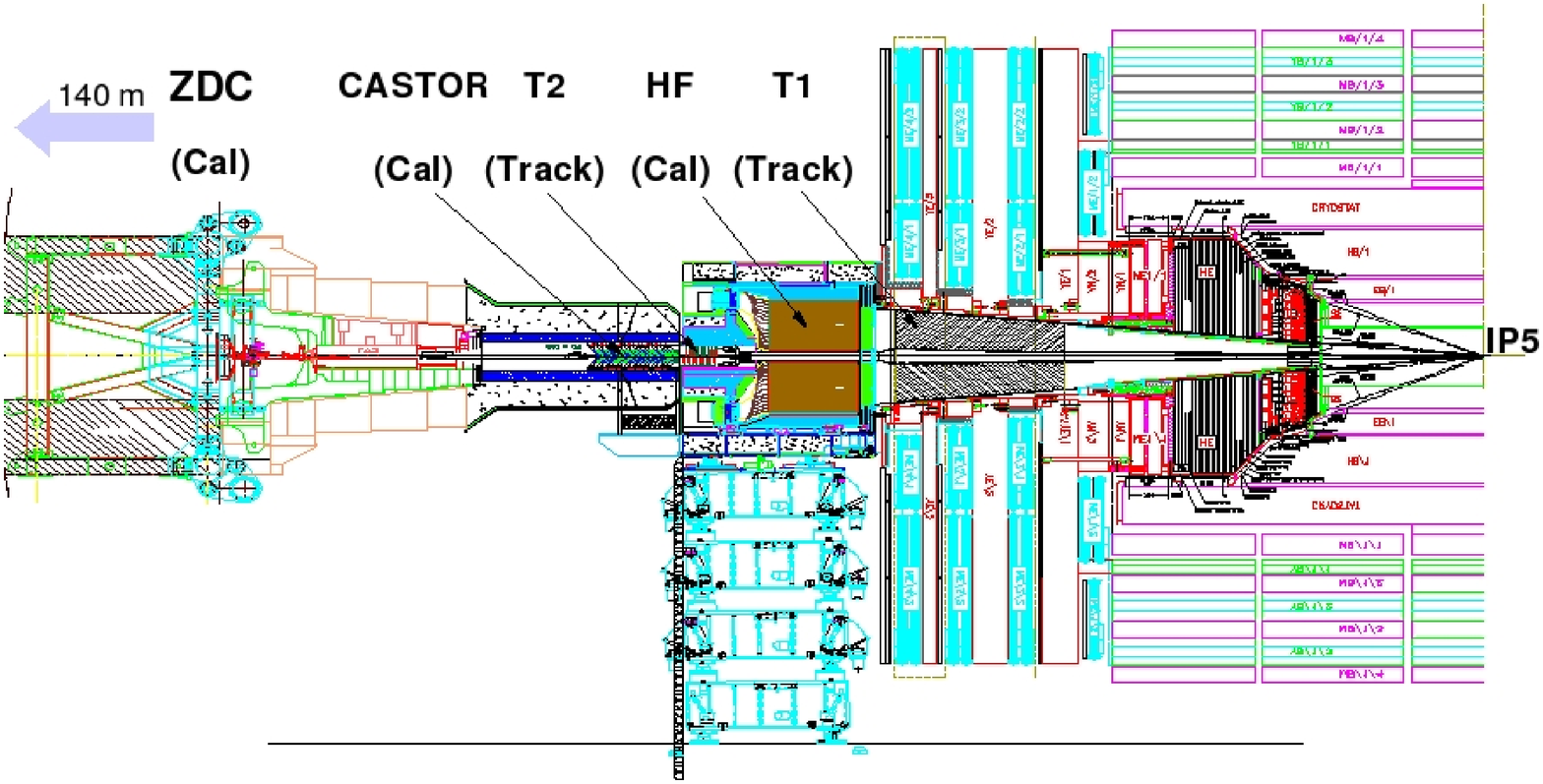}
\caption{Left: Log($x_{1,2}$) distribution of two partons producing at least one jet above $p_T$ = 20~GeV/c within HF and CASTOR.
Right: Layout of the detectors in the CMS forward region~\protect\cite{Fwd_LoI}. 
\label{fig:1}}
\end{figure}

In hadron-hadron collisions, the {\it minimum} momentum fractions probed in a  $2\rightarrow 2$ process with a 
particle of momentum $p_T$ produced at pseudo-rapidity $\eta$ are
\begin{equation}
x_{2}^{min} = \frac{x_T\,e^{-\eta}}{2-x_T\,e^{\eta}}\;\;,\;\;\;\; x_{1}^{min} = \frac{x_2\,x_T\,e^{\eta}}{2x_2-x_T\,e^{-\eta}}\;\;,
\mbox{ where } \;\; x_T=2p_T/\sqrt{s}\,,
\label{eq:x2_min}
\end{equation}
i.e. $x_2^{min}$ decreases by a factor of $\sim$10 every 2 units of rapidity. The extra $e^\eta$ 
lever-arm motivates the interest of {\it forward} particle production measurements to study low-$x$ QCD.
From Eq.~(\ref{eq:x2_min}), it follows that the measurement at the LHC of jets with transverse momentum $p_{T}$~=~20~GeV/c 
in the CMS forward calorimeters (HF, 3$<|\eta|<$5 and CASTOR, 5.1$<|\eta|<$6.6) will allow one 
to probe $x$ values as low as $x_{2}\approx 10^{-5}$. Figure~\ref{fig:1} (left) shows 
the log($x_{1,2}$) distribution for parton-parton scatterings in p-p collisions at 14~TeV producing at least 
one jet above 20 GeV/c in the HF and CASTOR acceptances. We present here simulation studies of two forward-jet 
measurements in CMS~\cite{CMS_FWD-2008/001} which are 
sensitive, respectively, to the low-$x_2$ (and high-$x_1$) proton PDFs
and to BFKL~\cite{mueller_navelet,DelDuca93,orr,sabiovera} or saturation~\cite{marquet,iancu08} QCD dynamics:
\begin{enumerate}
\item Single inclusive jet cross section in HF~\cite{hf} at moderate transverse momenta ($p_{T}\approx$ ~20 -- 120~GeV/c),
\item Differential cross sections and azimuthal (de)correlation of ``Mueller-Navelet''~\cite{mueller_navelet} 
dijet events, characterised by jets with similar $p_T$ separated by a large rapidity interval ($\Delta\eta\approx$~6 -- 10).
\end{enumerate}

The combination of HF, TOTEM, CASTOR and ZDC (Fig.~\ref{fig:1}, right) makes of CMS 
the largest acceptance experiment ever built at a collider. Very forward jets can be identified 
using the HF~\cite{hf} and CASTOR~\cite{castor} calorimeters. The HF, located 11.2 m away on both sides of the 
interaction point (IP), is a steel plus quartz-fiber \v{C}erenkov calorimeter segmented into 1200 towers of 
$\Delta\eta\times\Delta\phi\sim$~0.175$\times$0.175. It has $10 \lambda_I$ interaction lengths
and is sensitive to deposited electromagnetic (EM) and hadronic (HAD) energy.
CASTOR is an azimuthally symmetric EM/HAD calorimeter placed at 14.37 m from
the IP, covering 5.1$<|\eta|<$6.6. The calorimeter is a \v{C}erenkov-light device,
with successive layers of tungsten absorber and quartz plates as active medium
arranged in 2 EM (10 HAD) sections of about 22$X_0$ (10.3$\lambda_I$) radiation (interaction) lengths.

\section{Forward jets reconstruction in HF}
\label{sec:fwd_jets}
\begin{figure}[htb]
\includegraphics[width=0.49\textwidth,height=5.8cm]{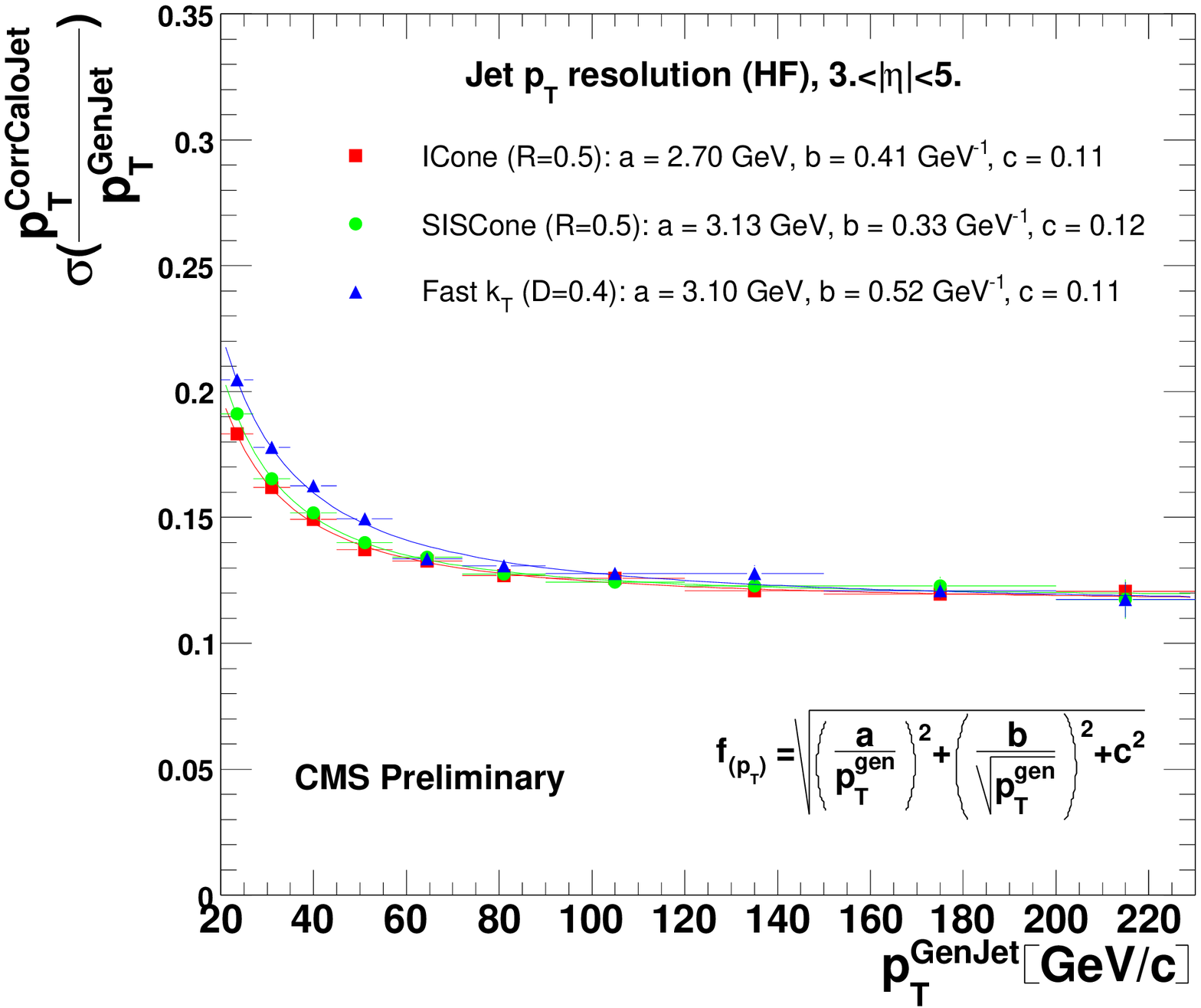}
\includegraphics[width=0.47\textwidth,height=5.7cm]{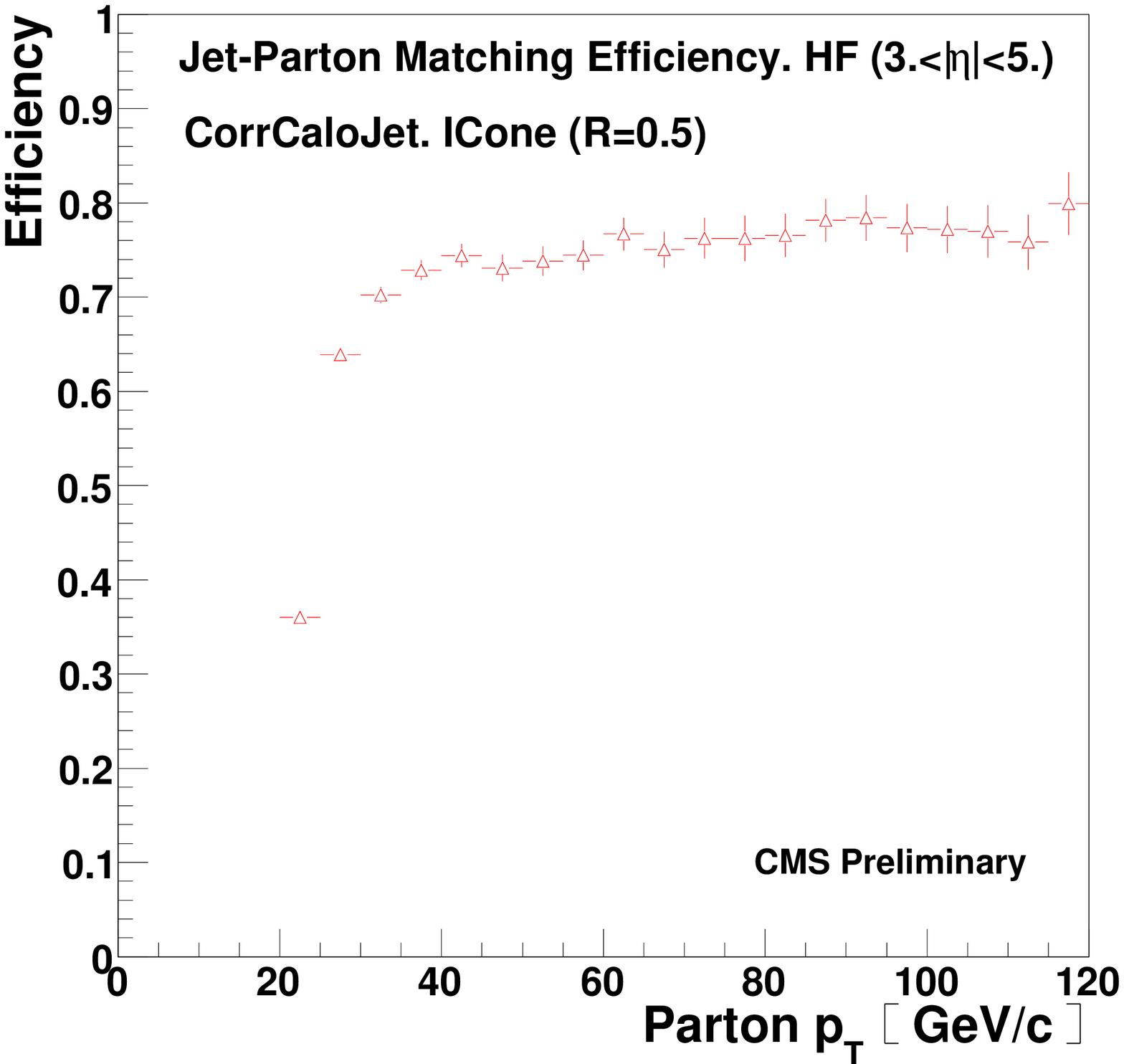}
\caption{Jet performances of the HF calorimeter. Left: Energy resolutions for various jet algorithms versus $p_{T}$. 
Right: Jet-parton matching efficiency as a function of $p^{parton}_{T}$ for a matching cone of $\Delta R=0.2$.
\label{fig:2}}
\end{figure}

Figure~\ref{fig:2} (left) shows the $p_{T}$ resolutions 
for forward jets ($3<|\eta|<5$) reconstructed using three different algorithms: iterative cone 
(ICone)~\cite{CMS_AN-2008/001} with $(\eta,\phi)$-radius of ${\cal R}=0.5$  
($E_{seed}$~=~3~GeV and $E_{thres}$ = 20~GeV), SISCone
~\cite{Salam_SISCone} (${\cal R}=0.5$), and the $k_T$ algorithm~\cite{Catani} ($D=0.4$) 
as implemented in the FastJet package~\cite{Cacciari_FastKt}. The forward jets reconstructed with the 
cone algorithms have a slightly better resolution than those obtained with the $k_T$ algorithm.
We find a $p_T$ resolution of $\sim$19\% at 20 GeV/c decreasing to  $\sim$10\% above 100 GeV/c. 
Such resolutions are {\it better} than at central rapidities
~\cite{ptdr1} because (i) the relevant variable for calorimetry resolutions is the {\it total} energy of the shower 
which for a given $p_T$ is always larger at forward than at central rapidities because of the forward boost; 
(ii) the jets are more collimated at forward rapidities and, thus, the ratio of jet-size/detector-granularity 
is more favourable. 
The position ($\eta$, $\phi$) resolutions (not shown here) for HF jets are also very good: 
$\sigma_{\phi,\eta}~\approx$~0.045 at $E_T$~=~20 GeV, improving to $\sigma_{\phi,\eta}\approx$~0.02 above 100 GeV.
Figure~\ref{fig:2} (right) shows 
the jet-parton matching efficiencies as a function of $p_{T}$ for reconstructed jets in HF. 
The matching efficiency saturates at around 80\% above $\sim$35~GeV/c,
which we take as our lowest reconstructible jet $p_T$ in HF.

\section{Single inclusive forward jet spectrum}
\label{sec:single_fwd_jets}
\begin{figure}[htbp]
\centering
\includegraphics[width=0.49\textwidth,height=6.cm]{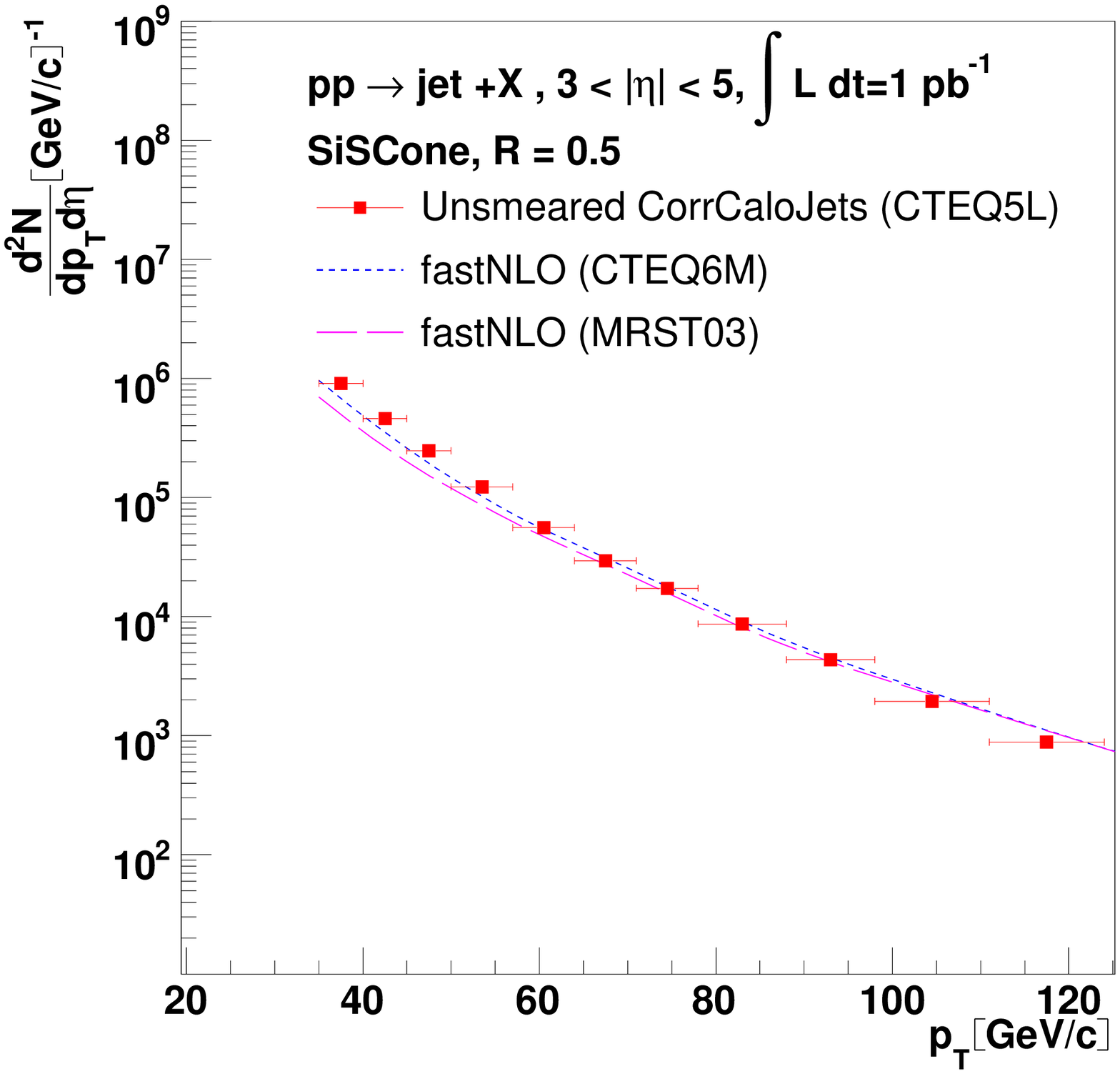}
\includegraphics[width=0.49\textwidth,height=6.cm]{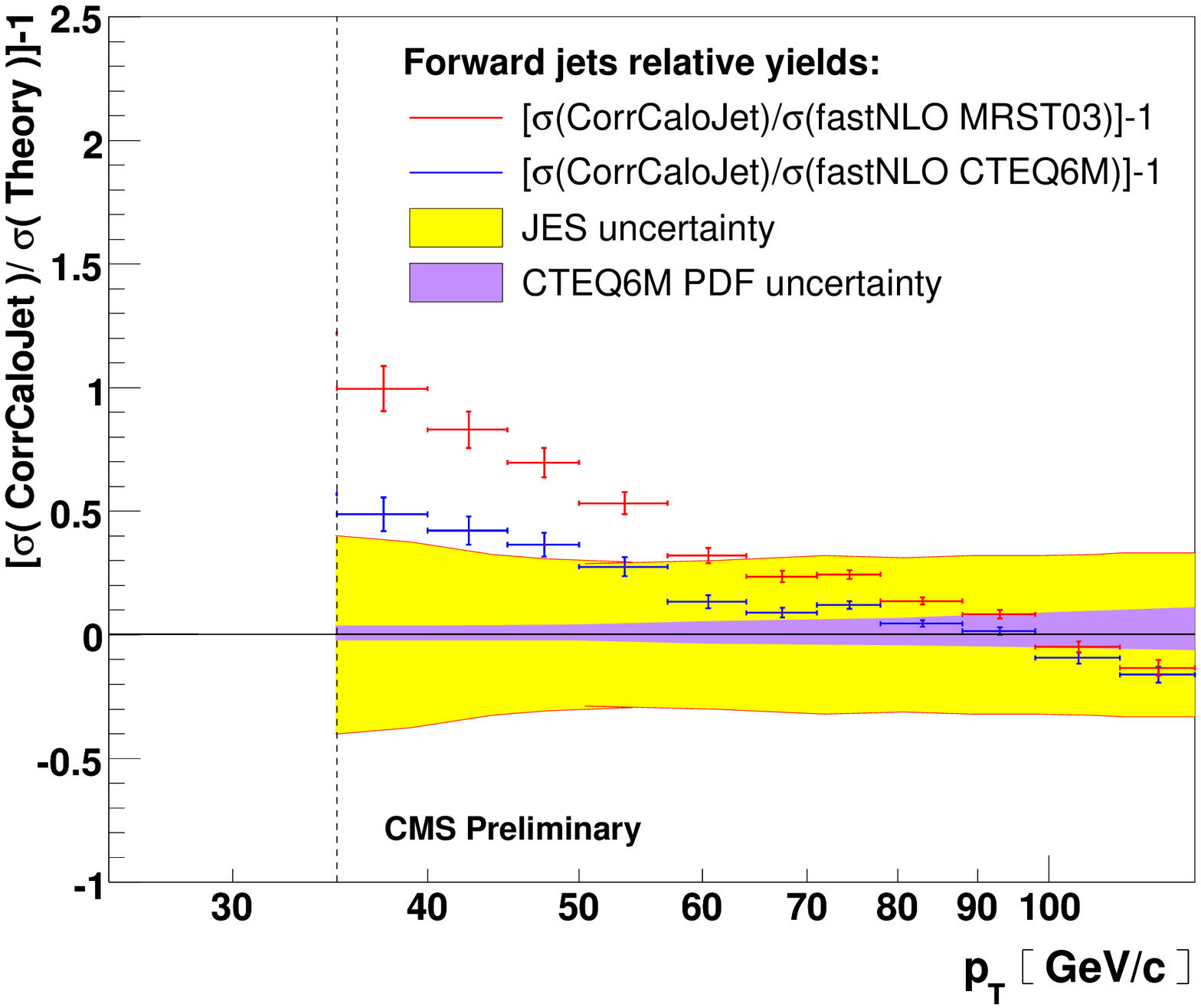}
\caption{Left: Forward jet yields for a total integrated luminosity of 1~pb$^{-1}$. 
Right: Percent differences between the reconstructed forward jet $p_T$ spectrum and fastNLO 
predictions for two PDFs~\protect\cite{fastnlo}.
\label{fig:3}}
\end{figure}

Figure~\ref{fig:3} (left) shows the expected jet $p_T$ spectrum in HF for p-p collisions at 14~TeV 
with an integrated luminosity of 1~pb$^{-1}$, reconstructed with the SISCone algorithm.
The spectrum has been corrected for energy resolution smearing but {\it not} for underlying-event and hadronization 
effects\footnote{Given the relatively small jet radius chosen, we expect those not to affect very significantly 
the result.}. The spectrum is compared to the fastNLO predictions\footnote{Renormalization-factorization scales: 
$\mu_i=p_T$. Other $\mu_i$ choices only change the spectrum by 5\%-9\%.}~\cite{fastnlo} 
for two different PDFs (MRST03 and CTEQ6.1M). The single jet spectra obtained with the two PDFs are similar at 
high $p_{T}$, while differences as large as $\mathcal{O}(60\%)$ appear below $\sim$60~GeV/c. 
Figure~\ref{fig:3} (right) shows the percent differences between the reconstructed spectrum and the 
two theoretical predictions. The error bars include the statistical and the energy-resolution smearing 
errors. The thin violet band around zero is the PDF uncertainty from the CTEQ6.1M set alone.
The main source of systematic uncertainty is due to the calibration of the jet energy-scale (JES). Assuming an early
10\% JES error, we find propagated uncertainties as large as 50\% in the jet yields at $p_T=$~40~GeV/c (yellow band)
which are similar to the theoretical uncertainty associated to the differences between PDF sets. 
If the JES can be improved at the 5\% or below, and the PDF uncertainties are indeed as large as the differences 
between MRST03 and CTEQ6.M, our forward jet measurement could help constrain the underlying 
PDF in global-fit analyses.
\vspace{-0.5cm}

\section{Mueller-Navelet dijets}
\label{sec:MN_dijet}
\begin{figure}[htb]
\centering
\includegraphics[width=0.49\textwidth,height=6.cm]{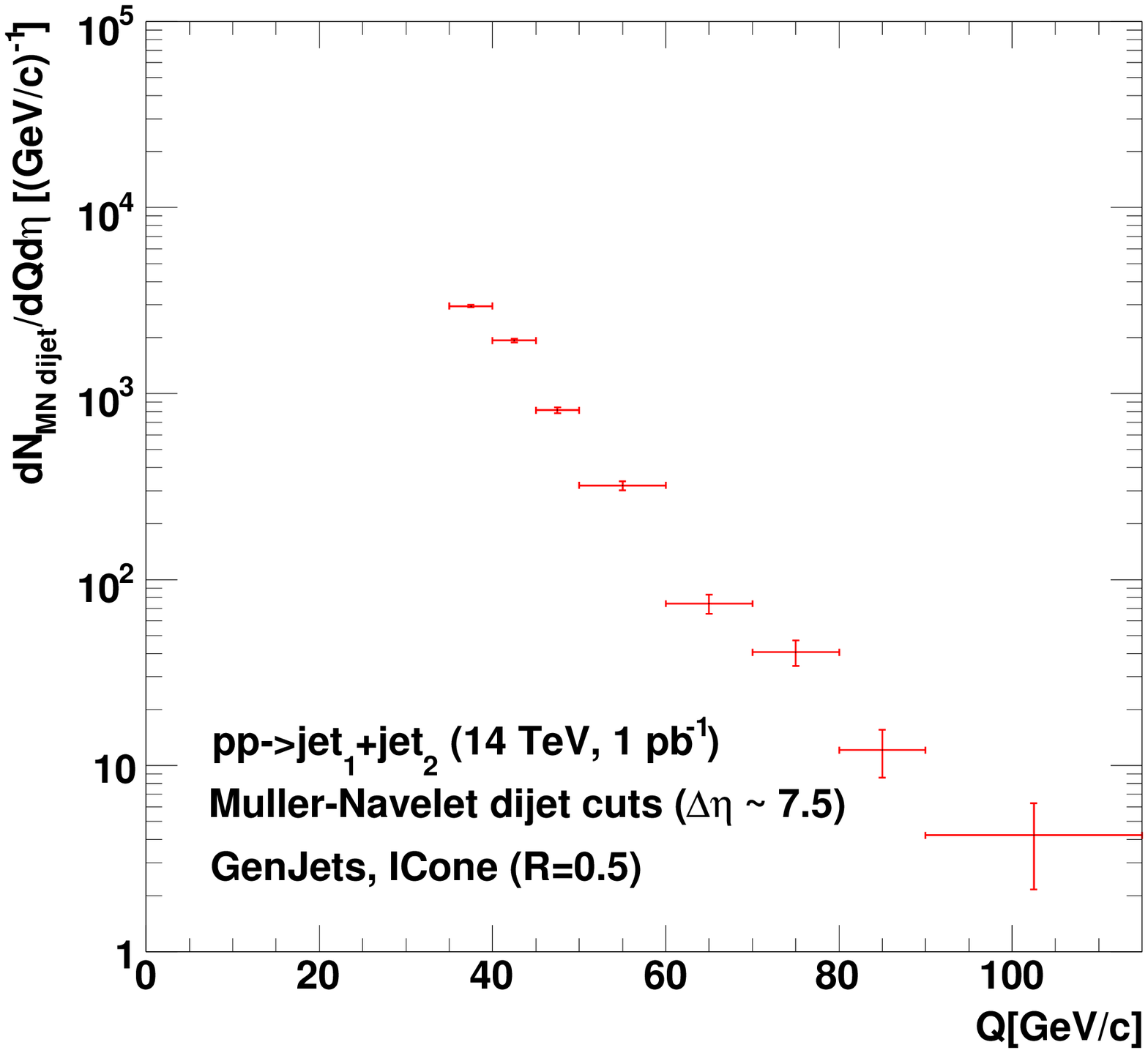}
\includegraphics[width=0.49\textwidth,height=6.cm]{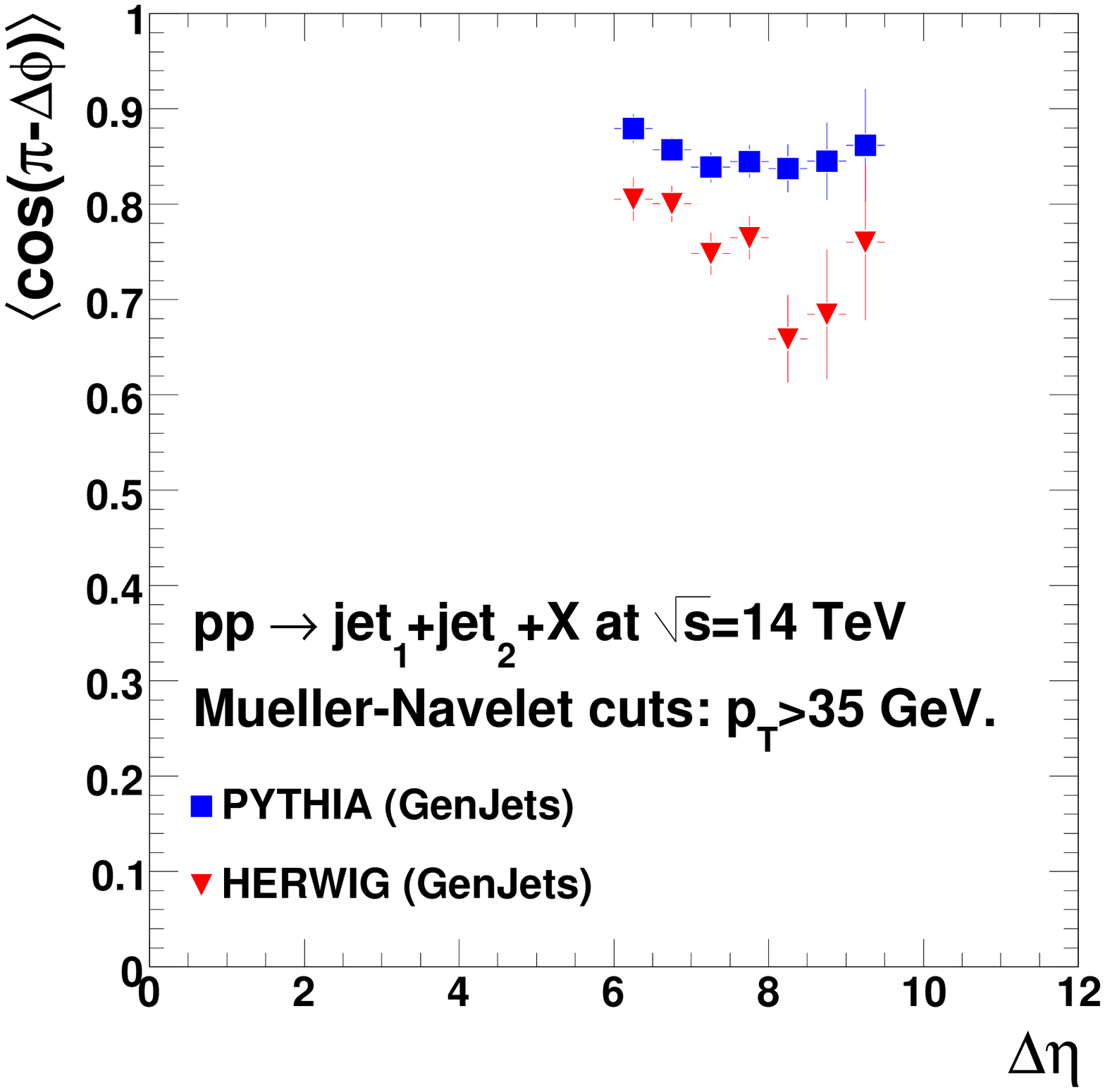}
\caption{Dijet events passing the Mueller-Navelet cuts (see text). 
Left: Expected \PYTHIA yields (1~pb$^{-1}$) for $\Delta\eta \approx$~7.5. 
Right: Average of $cos[(\pi - \Delta \phi)]$ versus $\Delta \eta$ in {\sc pythia} and {\sc herwig} dijet events.
\label{fig:muller_navelet_jets}}
\end{figure}

Mueller-Navelet (MN) events are characterized by two jets separated by a wide rapidity interval. 
Our experimental requirement to study such events is that each one of the jets is measured in a 
different HF calorimeter, i.e. that their separation is $\Delta\eta\gtrsim$~6. 
We generated events with~\PYTHIA~\cite{pythia6.4} and \HERWIG~\cite{herwig6}
and selected events with forward jets (ICone $\cal R =$~0.5) which satisfy the following Mueller-Navelet-type selection cuts:
\begin{itemize}
\item $p_{T,i} > 35$~GeV/c (good parton-jet matching and single-jet trigger efficiencies in HF)
\item $|\pTa - \pTb| < 5$~GeV/c (similar $p_{T}$ to minimise DGLAP evolution)
\item $3 <|\eta_{1,2}|< 5$ (both jets in HF) 
\item $\eta_1 \cdot \eta_2 < 0$  (each jet in a different HF)
\end{itemize}
The data passing the MN-cuts are divided into 4 equidistant HF pseudorapidity bins 
($[3.,3.5], ...,[4.5,5.0]$) and the dijet cross section  
computed as $d^2\sigma/d\eta dQ = N_{jets}/(\Delta\eta \Delta Q \int\mathcal{L}\mbox{dt})$, 
where $Q = p_{T,1} \approx p_{T,2}$ and  $N$ is the observed number of  jets in the $\Delta\eta, \Delta Q$ 
bin. For $\int\mathcal{L}\,\mbox{dt}$~=~1~pb$^{-1}$, 
we expect a few 1000s (100s) MN jets with separations $\Delta\eta>$~6~(9). Figure~\ref{fig:muller_navelet_jets}, 
shows the expected \PYTHIA yields passing the MN cuts for $\Delta\eta\approx$~7.5.
The obtained dijet sample appears large enough to carry out detailed studies of the $\Delta\eta$
dependence of the yields, and look e.g. for a possible ``geometric scaling'' in the Mueller-Navelet yields~\cite{iancu08}.\\

An enhanced azimuthal decorrelation for increasing rapidity separation between Mueller-Navelet 
jets is the classical ``smoking-gun'' of BFKL radiation~\cite{DelDuca93,orr,sabiovera,marquet}.
We have studied the dependence of the average value (over events) 
of the cosine of the $\Delta\phi$ difference between the jets, $\mean{cos\,(\pi - \Delta \phi)}$, 
as a function of their $\Delta\eta$ separation. One expects $\mean{cos(\pi - \Delta \phi)}$ = 1 (0) 
for perfect (de)correlation between the two jets.
The results are shown in Fig.~\ref{fig:muller_navelet_jets} (right)  
for the two highest-$p_T$ jets in the event passing the MN cuts. Only the dominant (statistical) errors are presented. 
At the Monte Carlo truth level (not shown here), the originating partons in \PYTHIA or \HERWIG are almost exactly back-to-back 
for all $\Delta\eta$ in each such jet-pair events. At the {\it generator-level}, the $\mean{cos(\pi - \Delta \phi)}$
decorrelation increases to 15\% (25\%) for \PYTHIA ({\sc herwig}), 
$\mean{cos(\pi - \Delta \phi)}\approx$~0.85~(0.75), 
due to parton showering and hadronization effects. 
Yet, the forward dijet decorrelation observed in both MCs is smaller (and less steep as a function of $\Delta\eta$) 
than found in BFKL approaches~\cite{sabiovera,marquet}.


\vspace{-0.2cm}

\begin{thebibliography}{999}
\bibitem{cgc}See e.g. F.~Gelis, T.~Lappi and R.~Venugopalan, \emph{Int.\ J.\ Mod.\ Phys. E}\ {\bf 16}, 2595 (2007) . 
\bibitem{dglap}V.N.~Gribov and L.N.~Lipatov, \emph{Sov.\ Journ.\ Nucl.\ Phys.}\ {\bf 15}, 438 (1972); 
G. Altarelli and G. Parisi, \emph {Nucl.\ Phys.}\,{\bf B126}, 298 (1977); 
Yu. L.~Dokshitzer, \emph{Sov.\ Phys.}\ JETP{\bf 46}, 641 (1977).
\bibitem{bfkl}L.N.~Lipatov, \emph{Sov.\ J.\ Nucl.\ Phys.}\, {\bf 23}, 338 (1976); 
E.A.~Kuraev,  L.N.~Lipatov and V.S.~Fadin, \emph{Zh. Eksp. Teor. Fiz} {\bf 72}, 3 (1977);  
Ya.Ya.~Balitsky, L.N.~Lipatov, \emph{Sov.\ J.\ Nucl.\ Phys.} {\bf 28}, 822 (1978).
\bibitem{CMS_FWD-2008/001}CMS Collaboration, CMS PAS FWD-08-001 (CMS-AN/2008-060) 
\bibitem{mueller_navelet}A.~H.~Mueller and H.~Navelet, \emph{Nucl.\ Phys.\ B} {\bf 282}, 727 (1987).  
\bibitem{DelDuca93}V.~Del Duca and C.~R.~Schmidt, \emph{Phys.\ Rev.\  D} {\bf 49}, 4510 (1994). 
\bibitem{orr}L.H. Orr, W.J. Stirling, \emph{Phys. Lett. B} {\bf 436}, 371 (1998). 
\bibitem{sabiovera}A.~Sabio~Vera, F.~Schwennsen, \emph{Nucl.\ Phys. B} {\bf 776}, 170 (2007); 
and private communication. 
\bibitem{marquet}C.~Marquet and C.~Royon,  \emph{Nucl.\ Phys.\ B} {\bf 739}, 131 (2006); 
and arXiv:0704.3409 [hep-ph]. 
\bibitem{iancu08}E.~Iancu, M.~S.~Kugeratski and D.~N.~Triantafyllopoulos, Nucl.\ Phys.\  A {\bf 808}, 95 (2008).
\bibitem{hf}A.~S.~Ayan {\it et al.}, \emph{J.\ Phys.\ G} {\bf 30}, N33 (2004). 
\bibitem{Fwd_LoI}M.~Albrow {\it et al.} [CMS and TOTEM Collaborations],  CERN/LHCC 2006-039/G-124 
\bibitem{castor}X.~Aslanoglou {\it et al.}, \emph{Eur.\ Phys.\ J.\  C} {\bf 52}, 495 (2007).  
\bibitem{CMS_AN-2008/001}CMS Collaboration, CMS PAS JME-07-003 (CMS/AN-2008-001)
\bibitem{Salam_SISCone}G.P.~Salam and G.~Soyez, {\it J. High Energy Phys.} 05 (2007) 086.
\bibitem{Catani}S. Catani, Y. L. Dokshitzer, M. H. Seymour and B. R. Weber, \emph{Nucl.\ Phys.\ B} {\bf 406}, 187-224 (1993).
\bibitem{Cacciari_FastKt}M.~Cacciari and G.P.~Salam, \emph{Phys.\ Lett.\ B} {\bf 641}, 57 (2006).
\bibitem{ptdr1}CMS Collaboration, CMS Physics TDR, Volume 1, CERN-LHCC-2006-001, 26 Feb. 2006. 
\bibitem{fastnlo}T.~Kluge, K.~Rabbertz and M.~Wobisch,  arXiv:hep-ph/0609285; 
and K.~Rabbertz, private comm.
\bibitem{pythia6.4} T.~Sjostrand, S.~Mrenna and P.~Skands, {\it J. High Energy Phys.} {\bf 0605} (2006) 026.
\bibitem{herwig6} 
  G.~Marchesini {\it et al.}, 
   \emph{Comput.\ Phys.\ Commun.}\  {\bf 67}, 465 (1992).  


\end{thebibliography}
\end{document}